\documentstyle[aps,prl,epsf]{revtex}
\def\roots{{\sqrt s}}
\newcommand{\mett}{\mbox{${\rm \not\! E}_{\rm T}$}}
\def \Et {{\rm E}_{\rm T}}
\def\Z{{ Z^0}}
\def\selectron{\tilde{e}}
\def\goes{\rightarrow}
\def\Gravitino{\tilde{G}}
\newcommand{\eeggmett}{ee\gamma\gamma\mett}
\def\pbarp{{\bar p}p}
\def\WWgg{WW\gamma\gamma}
\newcommand{\etal}{{\em et al.}}
\def\degrees{^\circ}
\def\ppbar{{\bar p}p}
 \newcommand{\NMETLOW              }{    1}
 \newcommand{\NMETHIGH             }{    2}
 \newcommand{\NIIIJETHIGH          }{    0}
 \newcommand{\NIVJETLOW            }{    2}
 \newcommand{\NGGTOT               }{ 2239}
 \newcommand{\NCENTEORMU           }{    3}
 \newcommand{\NCENTTAU             }{    1}
 \newcommand{\NBTAG                }{    2}
 \newcommand{\NADDGAMMA            }{    0}
 \newcommand{\NCENTLEP             }{    4}
 \newcommand{\NIIIJETEXPHIGH       }{\mbox{$     1.7\pm      1.5$}}
 \newcommand{\NIVJETEXPLOW         }{\mbox{$     1.6\pm      0.4$}}
 \newcommand{\NMETEXPLOW           }{\mbox{$     0.5\pm      0.1$}}
 \newcommand{\NMETEXPHIGH          }{\mbox{$     0.5\pm      0.1$}}
 \newcommand{\NTAUEXPLOW           }{\mbox{$     0.2\pm      0.1$}}
 \newcommand{\NBEXPLOW             }{\mbox{$     1.3\pm      0.7$}}
 \newcommand{\NADDGAMMAEXP         }{\mbox{$     0.1\pm      0.1$}}
 \newcommand{\NCENTEORMUEXP        }{\mbox{$     0.3\pm      0.1$}}
 \newcommand{\PURITYLOW            }{\mbox{$   15\pm    4$}}
 \newcommand{\RESSLOPE             }{\mbox{$0.043\pm 0.007$}}
 \newcommand{\RESOFFSET            }{\mbox{$2.66\pm 0.34$}}
 \newcommand{\EEGGMETTOTRATEPRL    }{\scinotn{  1}{ -6}}
 \newcommand{\WWGGHIGHRATEPRL      }{\scinotn{  8}{ -7}}
 \newcommand{\EEGFMPRL             }{\scinotn{  8}{ -8}}
 \newcommand{\EFAKEGGMETTOTRATEPRL }{\scinotn{  6}{ -8}}
 \newcommand{\EFGGMPRL             }{\scinotn{  5}{ -8}}
 \newcommand{\OVERLAPTOT           }{\scinotn{  8}{ -9}}
\newcommand{\mettsm}{\mbox{\scriptsize ${\rm \not\! E}_{\rm T}$}}
\newcommand{\mettgmh}{\mbox{${\rm \not\! E}_{\rm T}>35$~GeV}}
\newcommand{\mettx}{\mbox{${\rm \not\! E}_{\rm T}^x$}}
\newcommand{\ett}{${\rm E}_{\rm T}$}
\newcommand{\ptt}{${\rm P}_{\rm T}$}
\newcommand{\NONE}{\mbox{$N_1$}}
\newcommand{\NTWO}{\mbox{$N_2$}}
\newcommand{\CONE}{\mbox{$C_1$}}
\newcommand{\NTGNO}{\mbox{$\NTWO \rightarrow \gamma \NONE$}}
\newcommand{\tanbeta}{\tan\beta}
\def\Journal#1#2#3#4{{#1} {\bf #2}, #3 (#4)}
\def\PrePrint#1{\mbox{hep-ph/#1}}
\def\PRL{\em Phys. Rev. Lett.}
\def\PRD{{\em Phys. Rev.} D}
\newcommand{\scinotn}[2]{\mbox{${#1}\times 10^{#2}$}}
 \newcommand{\KANELOWACC           }{     5.4}
 \newcommand{\KANELOWXSEC          }{     1.1}
 \newcommand{\KANELOWNEXPTOT       }{     2.4}
 \newcommand{\PHOEFFIDLOW          }{   68\pm    3}
 \newcommand{\PHOEFFIDHIGH         }{   84\pm    4}
 \newcommand{\ZEESAMPLE            }{ 2663}
 \newcommand{\EEGGTOTMET           }{   55 \pm 7}
 \newcommand{\EEGGPT               }{   48 \pm 2}
 \newcommand{\EEGGMASS             }{  232 \pm 4}
 \newcommand{\EEGGMETMASS          }{  307 \pm 9}
 \newcommand{\EEMASS               }{  163 \pm 3}
 \newcommand{\MUMUGGMASS           }{   92 \pm 1}
 \newcommand{\EGGMASS              }{   91 \pm 2}
\begin{document}                                             

\begin{center}

%
%

{\bf \Large
Searches for New Physics in Diphoton Events in $p{\bar p}$ collisions at 
$\roots= 1.8$~TeV
}

\end{center}
\font\eightit=cmti8
\def\r#1{\ignorespaces $^{#1}$}
\hfilneg
\begin{sloppypar}
\noindent
F.~Abe,\r {17} H.~Akimoto,\r {39}
A.~Akopian,\r {31} M.~G.~Albrow,\r 7 A.~Amadon,\r 5 S.~R.~Amendolia,\r {27} 
D.~Amidei,\r {20} J.~Antos,\r {33} S.~Aota,\r {37}
G.~Apollinari,\r {31}
% ...
\end{sloppypar}
\vskip .026in
\begin{center}
(CDF Collaboration)
\end{center}

\vskip .026in
\begin{center}
\r 1  {\eightit Argonne National Laboratory, Argonne, Illinois 60439} \\
\r 2  {\eightit Istituto Nazionale di Fisica Nucleare, University of Bologna,
I-40127 Bologna, Italy} \\
\r 3  {\eightit Brandeis University, Waltham, Massachusetts 02254} \\
\r 4  {\eightit University of California at Los Angeles, Los 
Angeles, California  90024} \\  
\r 5  {\eightit University of Chicago, Chicago, Illinois 60637} \\
\r 6  {\eightit Duke University, Durham, North Carolina  27708} \\
\r 7  {\eightit Fermi National Accelerator Laboratory, Batavia, Illinois 
60510} \\
\r 8  {\eightit University of Florida, Gainesville, FL  32611} \\
\r 9  {\eightit Laboratori Nazionali di Frascati, Istituto Nazionale di Fisica
               Nucleare, I-00044 Frascati, Italy} \\
\r {10} {\eightit University of Geneva, CH-1211 Geneva 4, Switzerland} \\
\r {11} {\eightit Harvard University, Cambridge, Massachusetts 02138} \\
\r {12} {\eightit Hiroshima University, Higashi-Hiroshima 724, Japan} \\
\r {13} {\eightit University of Illinois, Urbana, Illinois 61801} \\
\r {14} {\eightit Institute of Particle Physics, McGill University, Montreal 
H3A 2T8, and University of Toronto,\\ Toronto M5S 1A7, Canada} \\
\r {15} {\eightit The Johns Hopkins University, Baltimore, Maryland 21218} \\
\r {16} {\eightit Institut f\"{u}r Experimentelle Kernphysik, 
Universit\"{a}t Karlsruhe, 76128 Karlsruhe, Germany} \\
\r {17} {\eightit National Laboratory for High Energy Physics (KEK), Tsukuba, 
Ibaraki 305, Japan} \\
\r {18} {\eightit Ernest Orlando Lawrence Berkeley National Laboratory, 
Berkeley, California 94720} \\
\r {19} {\eightit Massachusetts Institute of Technology, Cambridge,
Massachusetts  02139} \\   
\r {20} {\eightit University of Michigan, Ann Arbor, Michigan 48109} \\
\r {21} {\eightit Michigan State University, East Lansing, Michigan  48824} \\
\r {22} {\eightit University of New Mexico, Albuquerque, New Mexico 87131} \\
\r {23} {\eightit The Ohio State University, Columbus, OH 43210} \\
\r {24} {\eightit Osaka City University, Osaka 588, Japan} \\
\r {25} {\eightit Universita di Padova, Istituto Nazionale di Fisica 
          Nucleare, Sezione di Padova, I-35131 Padova, Italy} \\
\r {26} {\eightit University of Pennsylvania, Philadelphia, 
        Pennsylvania 19104} \\   
\r {27} {\eightit Istituto Nazionale di Fisica Nucleare, University and Scuola
               Normale Superiore of Pisa, I-56100 Pisa, Italy} \\
\r {28} {\eightit University of Pittsburgh, Pittsburgh, Pennsylvania 15260} \\
\r {29} {\eightit Purdue University, West Lafayette, Indiana 47907} \\
\r {30} {\eightit University of Rochester, Rochester, New York 14627} \\
\r {31} {\eightit Rockefeller University, New York, New York 10021} \\
\r {32} {\eightit Rutgers University, Piscataway, New Jersey 08855} \\
\r {33} {\eightit Academia Sinica, Taipei, Taiwan 11530, Republic of China} \\
\r {34} {\eightit Texas A\&M University, College Station, Texas 77843} \\
\r {35} {\eightit Texas Tech University, Lubbock, Texas 79409} \\
\r {36} {\eightit Istituto Nazionale di Fisica Nucleare, University of Trieste/
Udine, Italy} \\
\r {37} {\eightit University of Tsukuba, Tsukuba, Ibaraki 315, Japan} \\
\r {38} {\eightit Tufts University, Medford, Massachusetts 02155} \\
\r {39} {\eightit Waseda University, Tokyo 169, Japan} \\
\r {40} {\eightit University of Wisconsin, Madison, Wisconsin 53706} \\
\r {41} {\eightit Yale University, New Haven, Connecticut 06520} \\
\end{center}

\clearpage

\begin{abstract}

We have searched for anomalous production of missing 
E$_{\rm T}$ ($\mett$), jets, leptons ($e, \mu, \tau$), $b$-quarks, 
or  additional photons in events containing two isolated,
central  \mbox{($|\eta|<1.0$)} photons with 
\mbox{$\Et>12$~GeV}.  The results are consistent with standard model
expectations, with the possible exception of one event that has
in addition to
the two photons a central electron,  a 
high-E$_{\rm T}$ electromagnetic
cluster, and large  $\mett$.  We set 
limits using two specific SUSY 
scenarios for production of diphoton events with $\mett$. 

\end{abstract}
\begin{center}
\vspace*{0.2in}
\hspace*{1.0in} PACS numbers 13.85Rm, 13.85Qk, 14.80.-j,14.80.Ly
\vspace*{0.2in}
\end{center}
%
%
%
\clearpage
%
%
  In many models involving physics beyond the standard model (SM),  cascade
decays of heavy new particles generate $\gamma\gamma$  signatures involving
missing transverse energy ($\mett$), 
jets,  leptons, gauge bosons ($W$, $\Z$, $\gamma$), and
possibly $b$-quarks~\cite{interest}. 
For example, in supersymmetric models with a light gravitino,
pair-production of selectrons which decay via $\selectron\goes e\NONE\goes
e\gamma\Gravitino$ produces the $\gamma\gamma$ final state along with 
$\mett$ and electrons. 
In the data taken during 1993-1995, an `$\eeggmett$'
candidate event~\cite{Park} was recorded with the CDF Detector~\cite{detector}. 
We have performed
a systematic search for other anomalous $\gamma\gamma$ events
by examining events with two isolated,
central  \mbox{($|\eta|<1.0$)} photons with 
\mbox{$\Et>12$~GeV} which contain $\mett$, 
jets, leptons ($e, \mu, \tau$), $b$-quarks, 
or  additional photons~\cite{Future PRD}. This search is based on 
85$\pm$7~pb$^{-1}$ of data from $\pbarp$ 
collisions at $\roots = 1.8$~TeV collected with  the CDF detector.
In this Letter we describe the results of the search, including
the $\eeggmett$ candidate event, and set limits on two 
SUSY models that have arisen to explain it.

%
%

We briefly describe here the relevant aspects 
of the CDF detector. 
The magnetic spectrometer consists of tracking devices inside a
\mbox{3-m} diameter, 5-m long superconducting solenoidal magnet which operates 
at 1.4~T. A four-layer silicon microstrip vertex detector (SVX)~\cite{svxnim}
is used to identify $b$ hadron decays. A set of vertex time projection chambers
(VTX) surrounding the SVX  is used to find the $z$ position of the $\pbarp$
interaction (${\rm z_{vertex}}$).  The 3.5-m long central tracking chamber
(CTC) is used to measure the momenta of charged particles. The calorimeter,
constructed of projective electromagnetic and   hadronic towers,   is divided
into a central barrel which surrounds the solenoid coil (\mbox{$|\eta|<1.1$}),
`end-plugs' (\mbox{$1.1<|\eta|<2.4$}),   and  forward/backward modules
(\mbox{$2.4<|\eta|<4.2$}).   Wire chambers with cathode strip readout   give
2-dimensional  profiles of electromagnetic showers in the central  and plug 
regions (CES and PES systems, respectively).  A system of proportional wire
chambers (CPR) in front of the central electromagnetic calorimeters uses the
1-radiation-length-thick magnet coil as a `preradiator',   allowing 
photon/$\pi^0$ discrimination on a statistical basis by measuring the conversion
probability~\cite{photons}. Muons are identified with the central muon chambers,
situated outside the calorimeters in the region \mbox{$|\eta|<1.1$}.

%
%

	The data sample selection starts with events with two photon candidates 
identified by the three-level trigger~\cite{trigger}. At Level 1, events are
required to have two electromagnetic calorimeter trigger-towers~\cite{ttower}
with measured $\Et$ of more than 4~GeV.  At Level 2, we require
the logical `OR' of two triggers,  one optimized for good background rejection
at low $\Et$ and the other for high efficiency at high $\Et$. The  low-threshold
diphoton trigger requires two   electromagnetic clusters~\cite{EM Cluster}  with
\mbox{$\Et>10$~GeV} 
and an isolation requirement of less than 4~GeV in a 3-by-3 array
of trigger-towers around the cluster; the high-threshold (16~GeV) trigger  has
no isolation requirement.  Corresponding Level 3 triggers require  cluster
energies calculated with the offline photon algorithm~\cite{photons} to be above
the 10~GeV and 16~GeV thresholds. The low-threshold trigger also requires the
clusters be in a restricted fiducial region of the calorimeter~\cite{Fiducial}. 

We use the following selection criteria offline:
a)~two isolated~\cite{iso} 
central electromagnetic clusters with \mbox{\ett$>12$~GeV} (where the 10~GeV
trigger becomes \mbox{$>98\%$} efficient); 
b)~no tracks, or only one track with  \mbox{\ptt$<1$~GeV}, pointing at either 
cluster (to remove electrons or jets); 
c)~pulse height and shape in the CES consistent with a shower due to 
a photon (to remove $\pi^0$ backgrounds); 
d)~no other photon candidate
within the same 15$^0$ segment of the 
CES (to remove $\pi^0$ backgrounds)~\cite{Wgamma};
\mbox{e)~$|{\rm z_{vertex}}|<60$ cm} (to maintain the projective geometry of
the calorimeter); and 
f)~no energy out-of-time with the collision (to suppress 
cosmic rays)~\cite{ETOUT}. 
For
events in which both photon candidates have \mbox{\ett $>22$~GeV} 
(where the 16~GeV
trigger becomes \mbox{$>98\%$} efficient) 
the fiducial and isolation requirements are relaxed~\cite{Fiducial,iso}.
The final data set consists of \NGGTOT\  events.

%
%

	The efficiency for identifying an isolated photon is measured using 
electrons in a  control sample of $\ZEESAMPLE$ $\Z\rightarrow e^+e^-$   decays
to be $\PHOEFFIDLOW\%$ for the 12~GeV selection criteria, and $\PHOEFFIDHIGH\%$
for the 22~GeV selection criteria, in each case approximately flat in \ett. 
Photon backgrounds are measured using the shower shape in the CES system for
\mbox{\ett$<35$~GeV}, where the difference between a single $\gamma$ or $\pi^0 
\rightarrow \gamma \gamma$  can be resolved, and the conversion probability  in
the CPR for \mbox{\ett$>35$~GeV}~\cite{photons}. Since the purity of the
$\gamma\gamma$ sample
is of less importance to searches for rare
signatures than the  efficiency, we have chosen selection criteria to keep
the efficiency high; however these admit a substantial number of background
events. 
The fraction of events in the sample
which contain two prompt photons is measured to be $\PURITYLOW$\%.

%
%
We search the diphoton events for the presence of $\mett$, jets, 
electrons, muons, taus, $b$-quarks, and additional photons.    
To minimize the effect of 
fluctuations due to mismeasurement of jet energies,
we recalculate the
\mett\ by making jet energy corrections which take into account cracks between
detector components and nonlinear calorimeter response~\cite{jets,top}.  In
the $\gamma\gamma + \mett$ search we remove events which have a jet
with uncorrected \mbox{\ett$>10$~GeV} 
pointing within 10~degrees in azimuth of the
\mett. The resulting resolution on either the  $x$
or $y$ component of $\mett$, determined from a study of $\Z$ bosons, is 
well-parameterized by $\sigma(\mettx) = (\RESOFFSET\; {\rm~GeV}) + (\RESSLOPE)
\times  \Sigma{\rm E_{\rm T}}$, where $\Sigma{\rm E_{\rm T}}$ does not include
the ${\rm E_T}$ of either electron. The criteria for identifying jets 
(uncorrected \mbox{\ett$>10$~GeV} and \mbox{$|\eta|< 2.0$}), 
electrons, muons and $b$-jets are identical to those  used in
the top-quark discovery~\cite{top}. The tau selection is the same as used in the
study of $t{\bar t}$  decays into $e\tau$ and $\mu\tau$ final 
states~\cite{Marcus}. 
Any third photon is required to have \mbox{\ett$>25$~GeV} and to
pass the high-threshold selection criteria.

\begin{figure}
\epsfysize=8.5cm
\vspace*{-1cm}
\hspace*{10mm}
\epsfbox{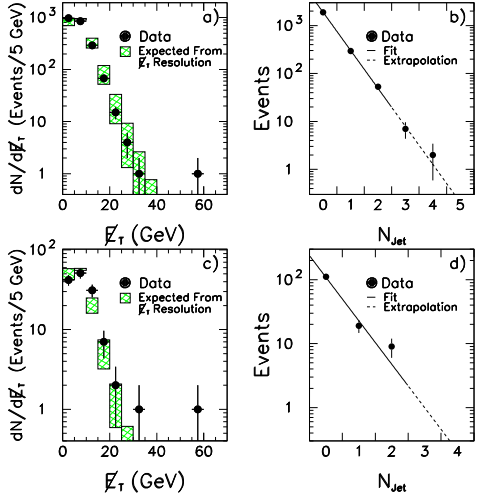}
\vspace*{-0.7cm}
\caption{
a) The $\mett$ spectrum for events with two  central
photons with  \mbox{\ett$>12$~GeV}. 
We have removed
events which have any jet with \mbox{\ett$>~10$~GeV} pointing 
within 10 degrees in azimuth
of the \mett.
The cross-hatched regions represent the background
estimates derived from the $\mett$ resolution in the $\Z$ control sample.
b) The spectrum in number of jets  with \mbox{\ett$>10$~GeV} and
$|\eta|< 2.0$ (N$_{\rm jet}$) 
for events with two central photons with
\mbox{\ett $>12$~GeV}.
c) and d) the same plots with photon \mbox{\ett$>25$~GeV}.
}
\label{Data Plots}
\end{figure}

Table~\ref{found}  summarizes the observed and expected numbers of events. The
distributions  in $\mett$ and the number of jets, N$_{\rm jet}$, are shown in  
Figure~\ref{Data Plots}.  
The shapes of the $\mett$ distributions are  in good agreement with the
resolution derived from  the  $\Z$ control sample, shown as the hatched region
in Figures~\ref{Data Plots}a and \ref{Data Plots}c. The distributions in  
N$_{\rm jet}$ are  well-modeled by an  exponential  extrapolation, shown in
Figures~\ref{Data Plots}b and \ref{Data Plots}d. For a photon threshold of
12~GeV we observe \NMETLOW\  event with \mbox{$\mett>35$~GeV}, 
with a SM expectation
of \NMETEXPLOW , and  \NIVJETLOW\ events with 4 or more jets,  versus an
expectation of \NIVJETEXPLOW.  For a photon  threshold of 25~GeV,   we observe
$\NMETHIGH$ events with \mbox{$\mett>25$~GeV}, with \NMETEXPHIGH\ expected, and 
$\NIIIJETHIGH$ events with 3 or more jets, with $\NIIIJETEXPHIGH$ expected.


%
%

We find \NBTAG\ events  with $b$-tags, consistent with background expectations,
and no events with a third photon. There are \NCENTLEP\ events with a central
lepton: one event is consistent with a double-radiative $\Z$ decay 
($m_{\mu\mu\gamma\gamma}=\MUMUGGMASS$~GeV/c$^2$), one is consistent with a
radiative $\Z$ decay with a lost track ($m_{ee\gamma}=\EGGMASS $~GeV/c$^2$), 
and one has a $\tau$ candidate, for which we expect a fake background of
\NTAUEXPLOW\ events.  From Table~\ref{found} and Figure~\ref{Data Plots}, we
find agreement between our observations and SM model predictions with one
possible exception.

The event that has the largest  
$\mett$ \mbox{($\mett=\EEGGTOTMET$~GeV)} among all diphoton candidates, 
has in addition to the  two
high-$\Et$ photons a central electron and an electromagnetic cluster in the plug
calorimeter which passes the electron selection criteria used for $\Z$ 
identification~\cite{R}.
The 4-vectors are presented in Table~\ref{Kinematics Table}. Because the 
momenta of the four clusters  are measured by the electromagnetic calorimeters,
the resolution on each is a few per cent (see Table~\ref{Kinematics Table}).  
The total \ptt\ of the 4-cluster system is 
$\EEGGPT$~GeV/c, opposite to the \mett\
and in good agreement with the measured magnitude, implying  the imbalance is
intrinsic to the 4-cluster system. 
The invariant mass of the electron and the
electromagnetic cluster  in the plug calorimeter is \mbox{$\EEMASS$~GeV/c$^2$},
far from the $\Z$ mass. The invariant mass of the 4-body system is
\mbox{$\EEGGMASS$~GeV/c$^2$}; a lower limit on the invariant mass of the total
system  is found by  including the $\mett$ (taking $p_z=0$), to be 
\mbox{$\EEGGMETMASS$~GeV/c$^2$}. %

%
%

Although the electromagnetic cluster  in the plug calorimeter passes all of the
standard electron selection criteria~\cite{R},  
there is no track in the SVX pointing directly at the cluster, as would be
expected if the cluster were due to an electron~\cite{Future PRD,Wasymm}. 
There is, however,  a track
26~mrad away  in $\phi$.  Using a sample of 1009 electrons in the end-plug
calorimeter  from $\Z\rightarrow e^+e^-$ events, we estimate the resolution on
$\phi$  to be 1.5~mrad; no events have a mismatch of greater than 20~mrad.  
The probability of an electron to have a  $\phi$ mismatch this large is thus
less than 0.3\% at 95\%~C.L.   The interpretation of the cluster as coming
from an isolated photon, the 1-prong hadronic decay of a $\tau$, or a jet, while
possible, are also all unlikely in that this would be an unusual example of any
of them~\cite{Future PRD}. 
We simply do not have enough  information to establish the origin of
the cluster.



%
%

	We have estimated the SM rates for producing a signature of  two
photons, two electromagnetic clusters (one central) passing the electron
requirements
and $\mett$, all with \mbox{$\Et>25$~GeV}, 
and \mbox{$m_{ee}>110$~GeV/c$^2$} (above the
$\Z$ boson)~\cite{Future PRD}. Using both data and Monte Carlo methods, we 
have considered production of SM
$WW\gamma\gamma$ and $t{\bar t}$, as well as sources which include additional
cosmic ray interactions, jets which fake electrons and/or photons, 
and overlapping events.
The total rate is \EEGGMETTOTRATEPRL\ events,  with
the dominant sources being $WW\gamma\gamma$ (\WWGGHIGHRATEPRL\ events) and 
$WW\gamma j$ (\EEGFMPRL\ events). Removing sources where the plug
cluster is due to a real electron, the rate is reduced to 
\EFAKEGGMETTOTRATEPRL\ events,  with the dominant source being 
$W\gamma\gamma j$ (\EFGGMPRL\ events).  Multiple events in the same beam
crossing~\cite{overlap} or the overlap of a cosmic ray interaction with a
$\ppbar$ event contribute a total of \OVERLAPTOT\ events. We
emphasize that while these SM estimates  are small and have led to valuable
speculation~\cite{interest},  it is indefensible to claim  evidence of new
physics based on one peculiar event selected from $3\times 10^{12}$ events.

%
%

One possible source of an $\eeggmett$ signature is anomalous $WW\gamma\gamma$
production. This hypothesis can be checked by searching for
events where each $W$ decayed hadronically rather than leptonically.
A Monte Carlo study using a standard model $\WWgg$ 
calculation~\cite{MrennaWWgg} shows that anomalous  
$\WWgg$ production would produce 
detected events with two photons with \mbox{\ett$>$25~GeV} and three 
or more jets 
30 times more often than events with two photons, two leptons and
\mett.  No events with three or more jets are seen in the 
N$_{\rm jet}$ distribution (Figure \ref{Data Plots}d).

%
%

We proceed to set limits on two SUSY 
models. There has been recent interest in supersymmetric models with either
the lightest neutralino (\NONE) 
decaying into a photon and
gravitino ($\Gravitino$) $[1b, 1c, 1e, 1g]$, 
\NONE $\goes \gamma \Gravitino$, or a supergravity scenario in
which the second- lightest neutralino decays via a loop into the lightest
neutralino and a photon~\cite{interest}c, 
\NTWO $\goes \gamma$\NONE. Both of these models
would produce events with two photons and $\mett$.

We 
use the SPYTHIA Monte Carlo~\cite{Spythia} with a full detector
simulation to  investigate the $\NTGNO$ model of  Ambrosanio
\etal\ with \mbox{M$_{\NONE}=36.6$~GeV}  
and \mbox{M$_{\NTWO}=64.6$~GeV}~\cite{Ambrosanio II}.
Direct production and cascade decays are predicted to produce 
$\KANELOWNEXPTOT$  events that pass the selection criteria of 
\mbox{E$_{\rm T}^{\gamma}>12$~GeV} and 
\mbox{$\mett>35 $~GeV}. In the data only the one event 
passes this selection;  we consequently cannot exclude
this model. To provide a normalization point ({\it e.g.} for  model-builders  to
estimate  the detector efficiency), we have simulated direct $N_2N_2$ 
production for  this same model and find an acceptance of $\KANELOWACC\%$.
Treating the one event as signal, and performing no background subtraction, 
we derive a 95\% C.L. cross  section upper limit of $\KANELOWXSEC$~pb.



\begin{figure}
\epsfysize=8.5cm
\vspace*{-1cm}
\hspace*{10mm}
\epsfbox{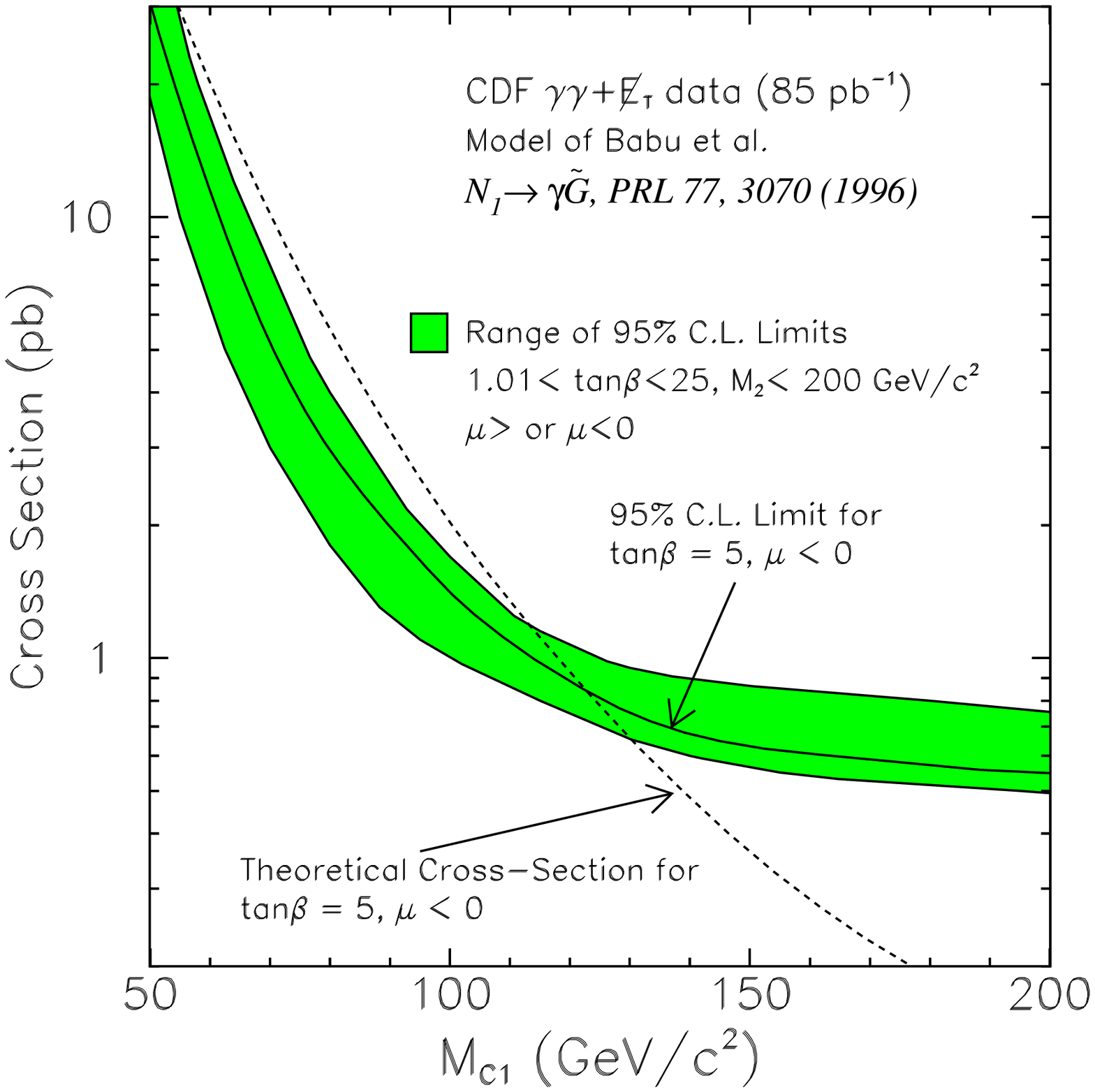}
\vspace*{-0.7cm}
\caption{The cross-section upper limits versus the mass of the $C_1$
for the light gravitino scenario of Babu \etal~\protect\cite{interest}e. The
shaded region shows the range of cross section limits as the parameters are
varied within the ranges
\mbox{$1<\tanbeta <25$}, 
\mbox{$M_2<200~$GeV}, and \mbox{$\mu>0$} or \mbox{$\mu<0$.}
The lines show the experimental limit and the theoretically predicted cross
section
for the lowest value of M$_{C_1}$ that is excluded
(\mbox{M$_{C_1}<120$~GeV} at 95\% C.L., for $\tanbeta=5$, 
\mbox{$\mu<0$}).}
\label{Limit Plots}
\end{figure}

Production of $\gamma\gamma$ events in the light gravitino scenario of Babu
\etal~\cite{interest}e 
is dominated by  $C_1N_2$ and $C_1C_1$  production and decay.
Figure~\ref{Limit Plots} shows the cross-section limits, using the same methods,
versus the mass of the \CONE.
The lines show the experimental limit and the theoretically predicted cross
section
for the lowest value of M$_{C_1}$ that is excluded
(\mbox{M$_{C_1}<120$~GeV} at 95\% C.L., for $\tanbeta =5$, \mbox{$\mu<0$}).
Note that because the $C_1$ and $N_1$ masses are related
we also exclude
\mbox{M$_{N_1}<65$~GeV} at 95\% C.L. \mbox{($\mu>0$}, \mbox{$\tanbeta=5$)}.
These limits are similar to those of the D$\O$
collaboration~\cite{D0}.

%
%

       	In  conclusion, we have  searched a  sample of 85  pb$^{-1}$ for events
with two central photons and anomalous production of missing transverse energy,
jets, 
charged leptons ($e, \mu$, and  $\tau$), $b$-quarks and photons. We find good
agreement  with  standard model   expectations, with the 
possible exception of one
event which has unusually large $\mett$ and in addition to the
two photons 
has a  high-$\Et$  central  electron  and a  high-$\Et$  electromagnetic
cluster. 


     We thank the Fermilab staff and the technical staffs of the
participating institutions for their contributions.  G.~Kane provided important
theoretical guidance. S.~Mrenna provided
critical help with SPYTHIA and with the $WW\gamma\gamma$ calculations. 
C.~Kolda provided invaluable assistance in the SUSY modeling.
We are also grateful to G.~Farrar, J.~Rosner, and 
F.~Wilczek for helpful conversations. This work was
supported by the U.S. Department of Energy and National Science Foundation,
the Italian Istituto Nazionale di Fisica Nucleare, the Ministry of Science,
Culture, and Education of Japan, 
the Natural Sciences and Engineering Research
Council of Canada, the National Science Council of the Republic of China; and
the A.P. Sloan Foundation.

\begin{table}
\begin{tabular}{lccc}
Signature (Object) & Obs. & Expected & Ref. \\
\hline
\mettgmh, $|\Delta\phi_{\mettsm-{\rm jet}}|>10^\circ$ 
                           & \NMETLOW      & \NMETEXPLOW           & -- \\
N$_{\rm jet}\ge 4$, ${\rm E}_{\rm T}^{\rm jet}>10$~GeV, 
$|\eta^{\rm jet}|<2.0$     & \NIVJETLOW    & \NIVJETEXPLOW    & \cite{top} \\
$b$-tag, ${\rm E}_{\rm T}^{b}>25$~GeV
                  & \NBTAG        & \NBEXPLOW             & \cite{top} \\
Central $\gamma$, ${\rm E}_{\rm T}^{\gamma_3}>25$~GeV
                        & \NADDGAMMA    & \NADDGAMMAEXP         & -- \\
Central $e$ or $\mu$, ${\rm E}_{\rm T}^{e~{\rm or}~\mu}>25$~GeV
    & \NCENTEORMU   & \NCENTEORMUEXP        &\cite{top}\\
Central $\tau$, ${\rm E}_{\rm T}^{\tau}>25$~GeV
          & \NCENTTAU     & \NTAUEXPLOW           &\cite{Marcus}\\
\end{tabular}
\caption{Number of observed and expected  $\gamma\gamma$ events with additional
objects in 85 pb$^{-1}$. The selection criteria, efficiencies, and background
estimation methods  used in identifying the jets, leptons and $b$-tags 
are discussed in the references.}
\label{found}
\end{table}

%
%
\begin{table}[th]
\begin{tabular}{lccccc}
  & P$_{\rm x} $  & P$_{\rm y}$  & P$_{\rm z}$ & E     & E$_{\rm T}$  \\
  & (GeV/c)       & (GeV/c)      & (GeV/c)     & (GeV) & (GeV)        \\
\hline
 $\gamma_1$   
   &     32.1(9)
   &    -16.8(5)
   &  -35(1)
   &   50(1)
   &   36(1)
 \\ \hline
 $\gamma_2$ 
   &    -12.9(4)
   &    -29.6(9)
   &    -22.5(7)
   &   39(1)
   &     32.3(9)
 \\ \hline
 $e^-$ 
   &  -34(1)
   &     11.5(3)
   &     21.7(6)
   &   42(1)
   &   36(1)
 \\ \hline
 Plug EM Cluster 
   &   60(2)
   &     19.0(5)
   & -172(5)
   &  183(5)
   &   63(2)
 \\ \hline
 \mett 
   &  -54(7)
   &   13(7)
  & ---
   &   ---
   &   55(7)
\end{tabular}
\caption{The 4-vectors of the objects in the $\eeggmett$ candidate event.
The parentheses represent the uncertainty in the last digit. There are no jets
with \mbox{E$_{\rm T}>10$~GeV}.}
\label{Kinematics Table}
\end{table}

%
%

\begin{thebibliography}{99}
\bibitem{Fiducial} 
The fiducial  region is  defined to be  0.05$<|\eta|<1.0$, 
\mbox{0$<\phi< 2\pi$}  with 86.7\% coverage for the high-threshold  trigger.  
For the low-threshold trigger,
the coverage is reduced to 73.3\%.
\bibitem{iso} 
For events in which either photon has \mbox{\ett$<22$~GeV}
the isolation requirement
is that a 3-by-3 array of trigger-towers around the photon cluster contain less
than 4~GeV. Isolation for events with both photons with 
\mbox{\ett$>~22$~GeV} 
is defined as the  ratio of the energy in a cone of 0.4 in
$\eta-\phi$ space, minus the photon cluster energy, over the cluster energy. We
require \mbox{Isolation$<$0.1}. 
In addition we require  that the sum of  the \ptt\ of
the tracks in the cone be less than 5~GeV, and  the  cluster energy in the
hadronic calorimeter divided by that in the electromagnetic be less than
$0.055+0.00045$E(GeV).
\bibitem{ETOUT} Only the central hadronic 
calorimeter has timing information associated with the energy deposited.  
We require there to be no towers with more than 1~GeV
deposited outside a window of $\pm 28$~nsec around the interaction time.
\bibitem{overlap} There are 4 primary vertices in the $\eeggmett$ candidate
event, versus  the mean expected number of 2.5.
\end{thebibliography}
\end{document}